\documentclass{PoS}

\def\br{\begin{eqnarray}}
\def\er{\end{eqnarray}}
\def\be{\begin{equation}}
\def\ee{\end{equation}}

\title{Chiral symmetry breaking with a confining propagator and dynamically massive gluons}

\ShortTitle{Chiral symmetry breaking with dynamically massive gluons}

\author{A. Doff \\
        Universidade Tecnol\'ogica Federal do Paran\'a - UTFPR,
Via do Conhecimento Km 01, 85503-390, Pato Branco - PR, Brazil\\
        E-mail: \email{agomes@utfpr.edu.br}}
        
\author{F. A. Machado, \speaker{A. A. Natale}\\
        Instituto de F\'{\i}sica Te\'orica - UNESP \\
        Rua Dr. Bento T. Ferraz,271 - Bloco II, 01140-070, S\~ao Paulo, SP, Brazil\\
        E-mail: \email{famachado@ift.unesp.br, natale@ift.unesp.br}}

\abstract{
Chiral symmetry breaking in QCD is studied introducing a confining effective propagator, as proposed recently
by Cornwall, and considering the effect of dynamically massive gluons. The effective confining propagator
has the form $1/(k^2+m^2)^2$ and we study the bifurcation equation finding limits on the parameter $m$ below which a satisfactory fermion mass
solution is generated. Since the coupling constant and gluon propagator are damped in the infrared, due to the presence
of a dynamical gluon mass, the major part of the chiral breaking is only due to the confining propagator and related to the low momentum region of the gap equation. We study the asymptotic behavior of the gap equation containing this confinement effect and massive gluon exchange, and find that the symmetry breaking can be approximated by an effective four-fermion interaction generated by the confining propagator. We compute
some QCD chiral parameters as a function of $m$, finding values compatible with the experimental data. 
We find a simple approximate relation between the fermion condensate and dynamical
mass for a given representation as a function of the parameters appearing in the effective confining propagator.} 

\FullConference{International Workshop on QCD Green's Functions, Confinement and Phenomenology,\\
		September 05-09, 2011\\
		Trento Italy}

\begin{document}

\section{Introduction}

QCD has two main properties: the chiral symmetry breaking (CSB) or dynamical quark mass generation and confinement of quarks and gluons.
Both phenomena are related to the non-perturbative infrared (IR) dynamics. The most usual non-perturbative
method to study these properties is QCD simulation on the lattice, but it is also possible to study dynamical mass generation in the
continuous space-time through Schwinger-Dyson equations (SDE) \cite{robwil}. Another fundamental QCD property is that the gluon may have
a dynamically generated mass, as suggested many years ago by Cornwall \cite{corna}, and this property has been thoroughly studied recently \cite{cornf,aguilara,corng}. We
may say nowadays that there are strong evidences for this fact as observed in lattice QCD simulations \cite{cucchi}, whose results
show nice compatibility with the SDE calculations \cite{aguilara}.

It is usually assumed that the fermionic gap equation can generate CSB only above a certain critical coupling 
equal to $\alpha_c (0)\equiv (g^2_c/4\pi) \geq (\pi/3C_2)$, where $C_2$ is the Casimir eigenvalue of the fundamental representation.
Actually it was suggested for fermions in a generic representation $R$ of non-Abelian gauge theories a 
hypothetical Casimir scaling law for fermion mass generation $\alpha_s C_2(R) \approx {\cal{O}}(1)$ \cite{susa},
which is a consequence of the fermionic SDE without the existence of a dynamical gauge boson mass.
Unfortunately the beautiful scenario of dynamical gluon mass generation, which may have deep implications for confinement \cite{corna}, poses a problem for the study of CSB through SDE. When gluons acquire a dynamical mass it has been shown that the coupling constant freezes in the
infrared, i.e. develops a non-perturbative infrared fixed point \cite{natale2a}, with the following behavior: 
\be
{\bar{g}}^2(k^2)= \frac{1}{b \ln[(k^2+4m_g^2)/\Lambda_{QCD}^2]} \, ,
\label{eq04}
\ee
where $b=(11N-2n_f)/48\pi^2$ for the $SU(N)$ group with $n_f$ flavors. For quarks in the
fundamental representation $C_2 = 4/3$ and $m_g\equiv m_g(k^2=0)\approx 2\Lambda_{QCD} \approx 500-600$ MeV, the phenomenologically preferred 
infrared value of the gluon mass \cite{natale},
this charge's value at the infrared fixed point ($\alpha_s (0)\equiv {\bar{g}}^2(0)/4\pi$) is of order $0.5$, while
it should be at least a factor $2$ larger to trigger CSB. Furthermore, the gluon propagator in the fermionic SDE kernel no longer behaves as $1/k^2$ but as $1/(k^2+m_g^2)$ in the infrared, what provides an extra damping in the gap equation. The consequence is that this equation does not generate dynamical quark masses ($M(k^2)$) compatible with the experimental data! However if quarks were in higher dimensional representation, with higher values for the Casimir operator, it would still be possible to generate some CSB \cite{cornwall2,aguilar2}.

The fact that if gluons acquire a dynamical mass we cannot obtain non-trivial solutions for the fermionic SDE was discussed in several papers \cite{haeria,haerib}. It could be said that the SDE are inappropriate to discuss CSB, but this is hardly the case in view of the successful results obtained through this approach in what concerns the gluonic sector. Therefore we can guess that some physical input is missing in the quark SDE. Before discussing a solution recently proposed by Cornwall to solve this problem \cite{cornwall3}, it is interesting to recollect some lattice and phenomenological results that are intimately
related to this problem. One important result of heavy quark phenomenology, consistent with lattice QCD, is that the potential felt by quarks in the fundamental representation is given by 
$V_F (r) = K_F r - \frac{4}{3} \frac{\alpha_s}{r}$,
where the first term is linear with the distance and proportional to the string tension $K_F$ generating a confining force. The second term, of order $\alpha_s$, the strong coupling constant, describes the one gluon exchange contribution. On the other hand the classical potential
between static quark charges is related to the Fourier transform of the time-time
component of the full gluon propagator in the following way
\be
V ({\bf{r}}) = - \frac{2C_2}{\pi} \int d^3 {\bf{q}} \alpha_s ({\bf{q}}^2) \Delta_{00}({\bf{q}}) \exp^{\imath {\bf{q.r}}} \,\, ,
\label{pot1}
\ee
where the bold terms, ${\bf{q}}$ and ${\bf{r}}$, are 3-vectors. As noticed in Ref.\cite{vento} the linear confining term of the potential ($K_F r$) cannot be obtained from the gluon propagator determined in the lattice or from the gluonic SDE, i.e. we could roughly say that the dynamically massive gluon propagator also does
not lead to quark confinement as it may not lead to CSB.

Another important QCD lattice result that is connected to the CSB mechanism, is the observation that most of the chiral breaking is related to the
very low momenta component of the gluon, i.e. CSB is associated to the deep infrared region \cite{sugaa}, which is the region dominated by the linear confining potential. Finally, from a phenomenological point of view, we cannot neglect the many successful results obtained by Nambu-Jona-Lasinio type of quark models, which are effective four-fermion interactions explaining most of the CSB strong interaction phenomenology. We believe that it
is extremely difficult to generate such effective theories when we deal only with massive one-gluon exchange, with their infrared damped propagator and frozen coupling constant. 

We can resume the previous paragraphs saying that lattice QCD appear to demand a linear potential between quarks and a CSB mechanism
typical of very low momenta gluons, and on the other hand the phenomenological data seems to demand a strong interaction between quarks in order to
generate an effective four-fermion interaction. We can add to these arguments the following results: a) $SU(2)$ lattice simulations have shown
that that the removal of confining center vortices restore the chiral symmetry \cite{bow1,holl1}; b)
In the case of adjoint fermions we may have CSB without confinement \cite{kar}, what may be due to the large Casimir value present in the
gap equation with massive one-gluon exchange \cite{cornwall2,aguilar2}. Therefore the one-gluon massive gap equation may drive CSB, although this breaking will be
related to the exchange of intermediate momenta gluons \cite{aguilar2}. We then have several indications that confinement may
play an important role in CSB. Cornwall proposed recently an effective propagator for the gap equation, based on confinement and entropy criteria, that cures the problem of CSB in the presence of dynamically massive gluons \cite{cornwall3}. We complement Cornwall's paper in
some points, and, particularly, show that most of the chiral breaking happens at very low momenta, what is compatible with the lattice simulations. We study the bifurcation equation finding limits on the parameters appearing in the effective propagator, below which a satisfactory fermion mass solution is generated. We study the asymptotic behavior of the new gap equation containing the confinement effect and massive gauge boson exchange, and find that the symmetry breaking can be approximated by an effective four-fermion interaction generated by the confining propagator, in agreement
with phenomenological quark models with this type of interaction. We compute some QCD chiral parameters, finding values compatible with the experimental data, and obtain a simple approximate relation between the fermion condensate and dynamical quark mass for a given representation as a function of the parameters appearing in the effective confining propagator.

\section{Confining propagator and bifurcation of the gap equation}

Confinement was introduced explicitly into the gap equation through the 
following {\it effective propagator}, {\it which is not at all related to the propagation of a standard quantum field} \cite{cornwall3}:
\be
D_{eff}^{\mu \nu}(k) \equiv \delta^{\mu \nu} D_{eff} (k); \,\,\,\,\,  D_{eff} (k)=\frac{8\pi K_F}{(k^2+m^2)^2}   \, .
\label{eq01}
\ee
In the $m\rightarrow 0$ limit we would obtain the standard effective propagator $8\pi K_F \delta^{\mu \nu}/k^4$, that yields approximately
an area law for the Wilson loop. We must necessarily have a finite $m\neq 0$ value due to entropic reasons as demonstrated in Ref.\cite{cornwall3}, and its value is related to the dynamical quark mass ($m\approx M(0)$), as required by gauge invariance, originating a negative term $-K_F/m$ 
in the static potential in order to generate the Goldstone bosons associated to the chiral symmetry breaking. Moreover, the Abelian gauge invariance of this effective propagator must appear in the quark action obtained by integrating over quark world lines that will
imply a area-law action \cite{cornwall3}. It is opportune to remember that for many years it was thought, due to the known Mandelstam approximation \cite{mand}, that a confining $1/k^4$ propagator would come out naturally from the Schwinger-Dyson equations. Now, after recognizing that the
gluon propagator has a massive behavior \cite{aguilara} and is a ``confined" gluon, it seems that we ought to change the paradigm and introduce confinement explicitly into the gap equation as proposed in Ref.\cite{cornwall3}.

The full gap equation in this case is given by:
\br
M(p^2)&=&\frac{1}{(2\pi)^4}\int \, d^4k \, D_{eff} (p-k) \frac{4M(k^2)}{k^2+M^2(k^2)}   \nonumber \\
&&+\frac{C_2 }{(2\pi)^4}\int \, d^4k \,  \frac{{\bar{g}}^2(p-k)3M(k^2)}{[(p-k)^2+m_g^2(p-k)][k^2+M^2(k^2)]} \, ,
\label{eq0511}
\er
where $M(p^2)$ is the dynamical quark mass generated by the confining and dressed-gluon propagator. Apart from the approximate
solutions of Eq.(\ref{eq0511}) found in Ref.\cite{cornwall3} we can study its critical behavior examining its bifurcation equation.
This equation is a standard Fredholm equation with a positive kernel, and, requiring its solution to belong to $L^2$, the 
spectrum is discrete with a smallest value for the ``effective coupling" $K_F/m^2$ and the $1$-gluon exchange
coupling ${\bar{g}}^2/4\pi$ such that we have the trivial solution for values of these couplings smaller than a certain critical value, and the nontrivial one if their values are larger than this same critical value.
If we linearize Eq.(\ref{eq0511}) and define the variables $x=p^2/M^2$, $y=k^2/M^2$, $\kappa = m^2/M^2$, $\epsilon = m_g^2/M^2$,
$\rho = \Lambda_{QCD}^2/M^2$ and $f(p^2)= \delta M (p^2)/M$, we obtain the following bifurcation equation \cite{dmn} 
\be
f(x)=\frac{1}{\pi}\int_0^{\Lambda^2/M^2} \, dy \, K(x,y) f(y) \,\, ,
\label{eq09}
\ee
where we introduced an ultraviolet cutoff ($\Lambda$) and the kernel $K(x,y)$ is equal to
\br
K(x,y)&=&  \frac{y}{(y+1)} \left[ \left( \frac{2K_f}{M^2}\frac{1}{(y+\kappa )^2} +\frac{3C_2}{16\pi} \right. \right.
\left. \frac{{\bar{g}}^2(y)}{(y+\epsilon)} \right) \theta (y-x) \nonumber \\
&+& \left( \frac{2K_f}{M^2}\frac{1}{(x+\kappa )^2} +\frac{3C_2}{16\pi} \right.
\left. \left. \frac{{\bar{g}}^2(x)}{(x+\epsilon)} \right) \theta (x-y)\right] \,\, .
\label{eq10}
\er
The kernel $K$ is square integrable and the first bifurcation point of the 
nonlinear equation must satisfy $\frac{1}{\pi} \left\| K \right\| = 1 $. 
This bifurcation condition lead us to Fig.(\ref{fig1}), whose curves were obtained in the case of $n_f =3$,
$\Lambda_{QCD} =300$MeV and $K_F = 0.18$GeV$^2$. The dashed (blue) curve was obtained with $m_g=600$MeV, the dot-dashed (red) curve
with $m_g=650$MeV and the solid (black) curve with $m_g=700$MeV. Each point of these curves indicates the bifurcation point for a given
$m$ value generating a dynamical quark mass $M$. It should be noticed that there is a maximum $m$ value above which there is no CSB.
It is also interesting 
to verify that CSB also receives contributions from the massive gluon term, and this is the reason for the differences between the curves,
though the massive gluon (or $1$-gluon exchange) generates a minute mass, as already observed many years ago in Ref.\cite{haeria,haerib}. As
the dynamical gluon mass is decreased we can observe a small increase in the maximum value of the $m$ parameter, due to the 
$1$-gluon exchange increasing contribution to the gap equation.
The confining effective propagator dominates the amount of symmetry breaking and has most of its effect concentrated in a momentum region 
below ${\cal{O}}(100)$ MeV. Actually, as we shall discuss ahead, most of the chiral symmetry breaking will occur due
to the low momentum region of the gap equation, what is consistent with lattice observations that the relevant
momentum component of gluons for CSB is exactly this region \cite{sugaa}. This fact may be contrasted with
the results of Ref.\cite{aguilar2}, where a larger part of the CSB comes from an intermediate region of ${\cal{O}}(1)$ GeV.
The details of our calculation can be found in Ref.\cite{dmn}.

%%%%%%%%%%%%%%%%%%%%%%%%Fig.1 %%%%%%%%%%%%%%%%%%%%%%%%%%%%%%%%%%%%%
%          plot  - Kernel
%%%%%%%%%%%%%%%%%%%%%%%%%%%%%%%%%%%%%%%%%%%%%%%%%%%%%%%%%%%%%%%%%%%
\begin{center}
{\bf \begin{figure}[ht]
\hspace*{3.8cm}\includegraphics[scale=0.8]{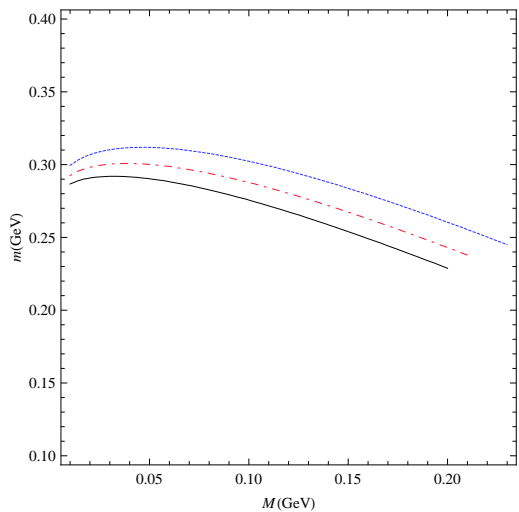}
\vspace{-0.1cm}
\caption{The bifurcation condition for the kernel of 
Eq.(2.4) 
is plotted in the case of $n_f =3$, $\Lambda_{QCD} =300$MeV and $K_F = 0.18$GeV$^2$. The dashed (blue) curve was obtained with $m_g=600$MeV, the dot-dashed (red) curve with $m_g=650$ MeV and the solid (black) curve with $m_g=700$MeV.}
\label{fig1}
%\vspace{-1.0cm}
\end{figure}}
\end{center}
%%%%%%%%%%%%%%%%%%%%%%%%%%%%%%%%%%%%%%%%%%%%%%%%%%%%%%%%%%%%%%%%%%% 

\section{The asymptotic behavior of the gap equation and the four-fermion approximation}

Eq.(\ref{eq0511}) can be transformed into a linear second-order differential equation with a singularity at infinity,
subjected to two boundary conditions. Its asymptotic
solution is a linear combination of two
independent solutions, $f(x)= b_1 f_+(x) + b_2 f_-(x)$, which can be obtained by applying the expansion method \cite{olver}.
In the ultraviolet limit we verify that the asymptotic behavior of the quark self-energy is, apart from logarithmic contributions, the known $1/p^2$ behavior (or $f_{reg}^{asymp}$ solution) found
by Politzer using the operator product expansion \cite{politzer}. Notice that the asymptotic behavior is fully described
by the $1$-gluon exchange, and this is not much different from what was found by Takeuchi \cite{takeuchi} in a problem where 
the CSB is dominated by a four-fermion interaction. The influence of the confining propagator enters only through the boundary 
conditions. As the mass solution that comes from the confining contribution has a fast falloff, almost behaving as an effective 
four-fermion interaction it is not surprising at all to see the similarity of our detailed results (see \cite{dmn}) with the ones of Ref.\cite{takeuchi}.
As discussed in Ref.\cite{dmn} for some peculiar values of the parameters appearing in the gap equation we may change the asymptotic 
behavior of the self-energy, from the regular behavior $f_{reg}^{asymp}$ to the one known as irregular ($f_{irreg}^{asymp}$), not differing from
the one with explicit CSB; this fact led us to study an actual four-fermion approximation to the gap equation. 

The gap equation has a different asymptotic behavior if the upper cutoff is of 
order $m\approx M$. 
Assuming an upper limit ($\Lambda$) in the momentum integration,
we studied in Ref.\cite{dmn} the ratio $R=f_{reg}^{asymp}/f_{irreg}^{asymp}$ when $\Lambda \approx m$ as a function of $a_1=\frac{2K_f}{\pi M^2}$ 
and $a_2=\frac{3C_2 {\bar{g}}^2}{16 \pi^2}$. This ratio is smaller than $1$ indicating that
the irregular asymptotic behavior of the symmetry breaking solution dominates over the regular one in this particular limit. 
In order to explore even more this possibility we can assume that the confining contribution could be reduced to an effective 
four-fermion interaction. Some of the reasons why we are concerned with the possibility of generating a four-fermion interaction are the following: 
First, confinement is introduced into the gap equation leading to a strong infrared force, we then
expect to reproduce some of the many phenomenological successful quark-models based on the Nambu-Jona-Lasinio type of interaction. Secondly,
lattice simulations show that the relevant gluonic energy scale of spontaneous CSB is due to the low-momentum component of the gluon field \cite{sugaa},
which may indicate the possibility of a natural upper cutoff in the momentum, appearing due to the saturation of the linear potential. The existence of a specific momentum that separates
the confinement and perturbative regions has also been discussed in a different context \cite{brodsky}.
Finally, the existence of a completely nonperturbative infrared fixed point, as happens when the theory develops a dynamical
gauge boson mass \cite{natale2a}, may induce effective four-fermion interactions as discussed many years ago in Ref.\cite{barda}.

It is known that 
as long as we have a massive gluon propagator it could be imagined that this mass could be factorized from
the propagator generating an effective four-fermion interaction, but this is not true because the actual interaction strength is
measured by the product ``coupling$\otimes$ propagator", and we know from Eq.(\ref{eq04}) that the $1$-gluon exchange has not enough
strength  to generate such effective interaction. On the other hand the confining effective propagator, with the usual values for
the string tension, is strong enough to generate the following effective gap equation:
\br
M_{4f}(p^2)&=&\frac{2}{\pi^3} \frac{K_F}{m^4} \int \, d^4k \,  \frac{M_{4f}(k^2)}{k^2+M_{4f}^2(k^2)} \theta (m^2-k^2) \nonumber \\
&+&\frac{C_2 }{(2\pi)^4}\int \, d^4k \,  \frac{{\bar{g}}^2(p-k)3M_{4f}(k^2)}{[(p-k)^2+m_g^2(p-k)][k^2+M_{4f}^2(k^2)]} \,\, .
\label{eq25}
\er
The solution of Eq.(\ref{eq25}) is similar to the one of 
Ref. \cite{takeuchi} observing the
interplay between its $4$-fermion coupling constant $\lambda$ and our effective coupling $K_F /m^2$.

In order to show how a $4$-fermion approximation is reasonable to describe the critical behavior of the complete Eq.(\ref{eq09}) studied
in the previous section, we can study the bifurcation problem for the $4$-fermion effective kernel given by 
\br
\left\| K_{4f} \right\|^2 &=& \int_0^\infty \, dx \, \int_0^x \, dy \,
 \frac{y^2}{(y+1)^2}\left(\frac{2K_f}{M_{4f}^2\kappa^2} \theta (\kappa -y)+ \frac{3C_2}{16\pi} \right.
\left. \frac{{\bar{g}}^2(x)}{(x+\epsilon)} \right)^2 \nonumber \\
&+& \int_0^\infty \, dx \, \int_x^\infty \, dy \, 
\frac{y^2}{(y+1)^2}\left(\frac{2K_f}{M_{4f}^2\kappa^2}\theta (\kappa -y) + \frac{3C_2}{16\pi} \right.
\left. \frac{{\bar{g}}^2(y)}{(y+\epsilon)} \right)^2 
\label{eq26}
\er

We separated the kernel of Eq.(\ref{eq26}) into two different kernels, one due to the effective four-fermion
interaction and another due to the exchange of a massive gluon. Using the triangle inequality 
$\left\| K_{4f} \right\| \equiv\left\| K_c +K_{1g} \right\| \leq \left\| K_c \right\| + \left\| K_{1g} \right\|$
and the bifurcation condition $\frac{1}{\pi} \left\| K_{4f} \right\| = 1$  we compute the critical condition and show in Fig.(\ref{fig3}) the dot-dashed (black) curve
of critical $m$ values for the generation of massive solutions of Eq.(\ref{eq25}). This curve was obtained for $m_g = 600$ MeV, $n_f =3$,
$\Lambda_{QCD} =300$ MeV and $K_F = 0.18$ GeV$^2$, and for comparison we also draw in Fig.(\ref{fig3}) the dashed (blue) critical curve of the
complete kernel given by Eq.(\ref{eq10}) (i.e. without the four-fermion approximation) 
computed with the same parameters. This shows that most of the symmetry breaking 
is driven by the confining effective
propagator and the four-fermion approximation is quite accurate up to an order of $10\%$, what is quite reasonable if we consider the
use of the triangle inequality to obtain Fig.(\ref{fig3}), indicating that the confinement effect, as proposed in
Ref.\cite{cornwall3}, may indeed generate an effective four-fermion interaction.

%%%%%%%%%%%%%%%%%%%%%%%%Fig.3 %%%%%%%%%%%%%%%%%%%%%%%%%%%%%%%%%%%%%
%          plot  - Kernel
%%%%%%%%%%%%%%%%%%%%%%%%%%%%%%%%%%%%%%%%%%%%%%%%%%%%%%%%%%%%%%%%%%%
\begin{center}
{\bf \begin{figure}[ht]
\hspace*{3.8cm}\includegraphics[scale=0.6]{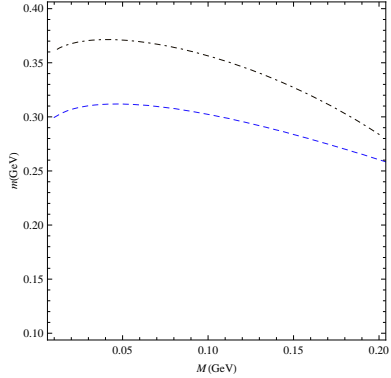}
\vspace{-0.1cm}
\caption{The bifurcation condition obtained for the kernel of 
Eq.(3.2) 
using the triangle inequality. The result for $m_g = 600$ MeV, $n_f =3$, $\Lambda_{QCD} =300$ MeV and $K_F = 0.18$ GeV$^2$ is shown by the dot-dashed (black) curve. For comparison we also draw the dashed (blue) critical curve of the complete kernel given by Eq.(2.4) 
computed with the same parameters.}
\label{fig3}
%\vspace{-1.0cm}
\end{figure}}
\end{center}
%%%%%%%%%%%%%%%%%%%%%%%%%%%%%%%%%%%%%%%%%%%%%%%%%%%%%%%%%%%%%%%%%%% 

Besides the fact that the gap equation generated in this approximation is numerically satisfactory, we
believe that the confining part of the fermionic SDE should have a natural upper cutoff at some scale not much
different from $m$. The reason for this is that the linear potential must break at some critical distance. For $n_f =2$ quarks in the fundamental representation, lattice QCD data shows that the string breaks at the critical distance $r_c \approx 1.25$ fm \cite{bali},
which corresponds to a $m$ value compatible with the one necessary for the expected amount of CSB. It would be outpurposed if confinement
(and, in particular, the confining propagator) were still effective to shorter distances or large momentum scales.
We can resume our results up to now (as detailed in \cite{dmn}): a) Most of
the CSB comes from a momentum region below ${\cal{O}}(100)$ MeV and b) The effective confining propagator seems to generate an
effective four-fermion interaction, which is not only observed through the bifurcation condition in Fig.(\ref{fig3}) but also through
the full numerical solution of the gap equation \cite{dmn}.

To confirm that the scenario of Ref. \cite{cornwall3} is fully consistent with the CSB phenomenology, we compute several
chiral parameters with the same values of $n_f$, $\Lambda_{QCD}$, $K_F$ and $m_g$ discussed above and $m=180$ MeV, which leads 
to the usually expected dynamical quark mass of $250$ MeV. These calculations are performed in the case of the full gap
equation (Eq.(\ref{eq0511})) and the results do not change appreciably from the values obtained with the four-fermion approximation. 
The chiral parameters, computed in the Abelian gluon approximation, that we consider are:
a) The pion decay constant $f_\pi^2 ={\bar{f}}_\pi^2 + \delta f_\pi^2$, where $\delta f_\pi^2$ is a correction
to ${\bar{f}}_\pi^2$ as determined in \cite{barducci}; b) The quark condensate $\left\langle {\bar{q}}q\right\rangle$ at the scale $\mu^2 = 1$ GeV$^2$; c) The MIT bag constant $\textsl{B}$.
The results for these three parameters, computed as a function of $m$, are shown in Table (\ref{tab1}). These values were calculated in the simple rainbow approximation \cite{dmn}, and better choices for the
vertex function as well as higher order corrections for these quantities can bring them closer to the experimental values. 

\begin{table}
\centerline{%
  \begin{tabular}{|l|c|c|c|c|c|}
    \hline \hline
    \quad $m$ \quad & \quad $\bar{f_{\pi}}$ \quad & \quad $f_{\pi}$\quad  &\quad  $\left\langle \bar{q}q\right\rangle(1\mbox{GeV}^2)$ \quad  & \quad  $B$ \quad \\ 
   \quad  ${[MeV]}$\quad  & \quad ${[MeV]}$\quad & \quad  ${[MeV]}$\quad  & \quad  ${[MeV^3]}$ \quad & \quad ${[MeV^4]}$ \quad \\ \hline 
    160 & 62.0 & 71.12 & 169.03 & 100.09\\ \hline
    180 & 54.51 & 61.83 & 156.12 & 84.96 \\ \hline
    200 & 46.59 & 52.14 & 142.0 & 69.43 \\ \hline
    Expected Values& 93& 93 & $229\pm 9$ &  146  \\ \hline \hline
  \end{tabular}}
\caption{Values of $f_\pi$, $\left\langle {\bar{q}}q\right\rangle$ and $\textsl{B}$ as a function of $m$ obtained with the numerical solution of Eq.(2.2).}
\label{tab1}  
\end{table}

\section{CSB for higher dimensional representations}

The study of CSB for fermions in higher dimensional representations is of interest because it is a possible way
to verify how this mechanism is distinct from the confinement one, as well as it is important for technicolor model building. 
If a type of Casimir scaling occurs \cite{susa}, we expect that for higher dimensional representations the CSB typical mass scale would
be different from the one for the fundamental representation, and perhaps different from the confinement scale.
It has also been argued that for ``quarks" in the adjoint representation the dynamically massive gluons may have enough
strength to generate a dynamical quark mass \cite{cornwall2,aguilar2}. Indeed in Ref.\cite{aguilar2} a large dynamical
mass was found for fermions in the adjoint representation, and one naively would expect that in this case the confining
and chiral breaking transitions would appear separately. 

If we follow straightforwardly the model
of Ref.\cite{cornwall3} we must also verify what is the difference introduced by the confining propagator in
the case of higher dimensional representations, because in principle we should replace the string tension $K_F$
by $K_R$, which is the string tension for fermions in the representation $R$. We assume that this replacement is
accurate, although we know that the phenomenological potential, and consequently the string
tension, does change according to the representation. For instance, in the case of the adjoint representation
it is known that $V_A (r \rightarrow \infty ) = 2 M_g $,
where $M_g$ is the energy of the lightest glueball. Moreover, the adjoint representation is not confined but
screened \cite{greensite}, what means that the confining propagator should be understood as effective up to
a certain distance in these cases. Of course, no matter what the fermionic representation is we shall have a
critical distance above which the string breaks. We assume that in the model of Ref.\cite{cornwall3} most of the chiral 
symmetry breaking is still related to the form of Eq.(\ref{eq01}), which does not get the chance to be probed at
large distances, consequently we may still expect that most of
the CSB is driven by the ``confining" propagator. 

The fermion condensate is the most frequent
quantity used to characterize the chiral phase transition, and it is this quantity that we will analyze to
investigate CSB for fermions in different representations.
It is easier to compute this quantity for different fermion
representations if we consider the gap equation in the four-fermion approximation given by Eq.(\ref{eq25}),
perform the angle approximation and neglect the gluon mass in the propagator of the $1$-gluon exchange contribution:
\br
M_{4f}(p^2)&=&\frac{2}{\pi^3} \frac{K_R}{m^4} \int_0^{m^2} \, d^4k \,  \frac{M_{4f}(k^2)}{k^2+M_{4f}^2(k^2)}  \nonumber \\
&+&\frac{C_2 }{(2\pi)^4}\int \, d^4k \,  \frac{{\bar{g}}^2(p)3M_{4f}(k^2)}{p^2[k^2+M_{4f}^2(k^2)]}\theta (p^2-k^2) \nonumber \\
&+&\frac{C_2 }{(2\pi)^4}\int \, d^4k \,  \frac{{\bar{g}}^2(k)3M_{4f}(k^2)}{k^2[k^2+M_{4f}^2(k^2)]} \theta (k^2-p^2) \,\, .
\label{eq54}
\er
On the other hand we may write the fermion condensate for the representation $R$ in the following form
\be
\left\langle {\bar{q}}q\right\rangle_R (m^2) = - \frac{N_R}{4\pi^2} \int_0^{m^2} \, dx \, \frac{x M_R(x)}{[x+M_R^2 (x)]} \,\, ,
\label{eq55}
\ee
where $N_R$ is the dimension of the fermion representation $R$ and $M_R(x)$ its dynamical mass. 
We are forcing the upper limit of Eq.(\ref{eq55}) to be of ${\cal{O}}(m)$, which
can be as low as $0.18$ GeV, while $\left\langle {\bar{q}}q\right\rangle$ is well known at the $1$ GeV scale.
We did this, as we shall see below, in order to easily compare the condensate expression to Eq.(\ref{eq54}), but
we can check that Eq.(\ref{eq55}) provides a good estimate of the quark condensate at $1$ GeV \cite{dmn}. 
We stress that the upper limit in the first integral in the right-hand side of Eq.(\ref{eq54}) may be a physical
one in order to be consistent with the critical distance at which the string breaks.

Eq.(\ref{eq55}) can be compared to Eq.(\ref{eq54}), the dynamical fermion mass $M_R$ in the four-fermion approximation, 
if we set all integrals at the scale $m$, obtaining
\be
M_R(m^2) \approx \left[ \frac{2K_R}{\pi m^4} + \frac{3C_{2R}g^2_R(m^2)}{16\pi^2 m^2}\right] \int_0^{m^2} \, dx \, \frac{xM_R(x)}{x+M^2_R(x)} \,\, .
\label{eq56}
\ee
Since the first term between brackets in the right-hand side of Eq.(\ref{eq56}) is much larger than the second one, combining the
two last equations we have $\left\langle {\bar{q}}q\right\rangle_R (m^2) \approx - \frac{N_R}{8\pi} \frac{m^4}{K_R} M_R(m^2) $.
For quarks in the fundamental
representation, this relation underestimates the condensate due to the fact that the integration area in this
equation was drastically reduced when we cutoff the integral at $m^2$. Since the effect of the Eq.(\ref{eq01}) is the dominating one, it
is quite plausible that the relation between the condensates holds up to other scales (still keeping the factor $m^4$ in
the right-hand side) and it can be tested through lattice simulations. 

We can now make a few comments on the differences between CSB for fermions in the fundamental and adjoint representations estimating the ratio of the condensates for $SU(3)$ fermions in these representations. First
we need to know how the string tension changes as we change the fermion representation.
It has been observed in lattice simulations what is usually called Casimir scaling for the string tension \cite{greensite}, i.e.
$K_R \approx \frac{C_R}{C_F} K_F $
where $C_R/C_F$ is the ratio between the Casimir operators for the representation $R$ and the fundamental one.
For $SU(N)$ theories and a finite $N$ the Casimir scaling law must break down at some point, and should be replaced by a dependence
on the $N$-ality $k$ of the representation $K_R = f(k) K_F$ \cite{greensite}.
This change of behavior is credited to an effect of force screening by the gluons. For fermions in the adjoint representation
the $N$-ality is zero, therefore, according to Casimir scaling, the adjoint string tension is given by
$K_A = \frac{2N^2}{N^2-1} K_F $,
and, as a reasonable approximation, we may assume $K_A \approx 2K_F$. 
Consequently we obtain the following ratio at the scale $m^2$
\be
\frac{\left\langle {\bar{q}}q\right\rangle_3 }{\left\langle {\bar{q}}q\right\rangle_8 } \approx \frac{3}{4} \frac{M_3}{M_8} \,\, .
\label{eq58}
\ee
Once the dynamical masses almost scale with the string tension value we could say that the above ratio is roughly of order $3/8$.
Of course, the uncertainty in this estimative is certainly connected to the remarks made at the beginning of this section about
the phenomenological potential and the effective propagator for the adjoint representation. For other fermionic representations
the screening behavior is smaller, although in all cases we certainly have a limit on the critical distance for which this approach
is valid that will be connected with the string breaking mechanism.  

\section{Conclusions}

We discussed chiral symmetry breaking in QCD and in the
presence of dynamically massive gluons. Confinement was introduced into the SDE in the form of an effective propagator,
which is one that can reproduce an area law for quarks. 
This is a phenomenological way to investigate the possibility, indicated by the lattice, of confinement (by center vortices)
being intrinsically related to chiral symmetry breaking.
We briefly review the conditions for the confining propagator to generate non-trivial massive solutions. We studied the bifurcation
condition for the complete gap equation, i.e. the one with the exchange of dynamically massive gluons and with the
inclusion of the confining effective propagator dependent on the entropic parameter $m$, which must be proportional
to the dynamical quark mass, and on the string tension $K_F$, verifying that there is a maximum $m$ value
below which the chiral symmetry is broken, and generating the expected values for the dynamical quark masses. As already known in
the literature we verified that the massive gluon exchange gives only a minute contribution to the dynamical quark mass.
Most of the breaking is due to the confining propagator and this may be one indication that we should not expect large differences
between the chiral and confinement transitions. 

It is known 
that the quark gap equation with a massive gluon cannot be reduced to an effective four-fermion interaction due to the small infrared strength
of the product ``coupling$\otimes$propagator", however we verified that this is not the case for the confining effective propagator. The
bifurcation equation for the gap equation with the four-fermion approximation performed in the confining part reproduces
approximately the result of the complete gap equation. Our numerical calculation of the quark mass
is consistent with the approximate results obtained by Cornwall \cite{cornwall3}. 

We discuss CSB for fermionic representations different than the fundamental one.
We found a simple relation between the fermion condensate and the fermionic dynamical mass. This relation depends on the
dimension of the fermion representation, $m$ and $K_F$, and we assumed that the form of Eq.(\ref{eq01}) still holds for
different fermionic representations up to a certain distance. In principle such relation can be studied in lattice simulations. 

Our results indicate that the CSB mechanism proposed in Ref.\cite{cornwall3} can account for the expected values of several
known chiral parameters. The model also seems to indicate that the CSB scale is connected to the confinement one even
for fermions in higher dimensional representations.
How far can we assume the four-fermion approximation discussed here for the purpose of practical 
calculations still needs further analysis. Finally, a more precise determination of the CSB observables can be obtained with 
the introduction of more sophisticated vertex functions and higher order corrections. 

\acknowledgments

We are indebted to A. C. Aguilar and  Prof. J. M. Cornwall for discussions. This research was partially supported by 
the Conselho Nacional de Desenvolvimento Cient\'{\i}fico e Tecnol\'ogico (CNPq) (AD and AAN), and Funda\c c\~ao de Amparo a Pesquisa do Estado de S\~ao Paulo (FAPESP) (FAM).

\end{document}